\documentclass[twocolumn,english,prl, showpacs, superscriptaddress]{revtex4-1}
\usepackage[T1]{fontenc}
\usepackage[utf8]{inputenc}
\usepackage{color}
\usepackage{babel}
\usepackage{float}
\usepackage{array}
\usepackage{booktabs}
\usepackage{multirow}
\usepackage{amsmath}
\usepackage{graphicx}
\usepackage[unicode=true,pdfusetitle,
 bookmarks=true,bookmarksnumbered=false,bookmarksopen=false,
 breaklinks=false,pdfborder={0 0 1},backref=false,colorlinks=false]
 {hyperref}

\makeatletter

\providecommand{\tabularnewline}{\\}

\hypersetup{
    colorlinks=true,
    linkcolor=red,
    citecolor=blue,
    filecolor=blue,      
    urlcolor=blue,
}
\makeatother

\begin{document}
\title{\textcolor{black}{Cooling with fermionic reservoir}}
\author{Gabriella G. Damas}
\address{Instituto de Fí­sica, Universidade Federal de Goiás, 74.001-970, Goiânia
- GO, Brazil}
\author{Rogério J. de Assis}
\address{Instituto de Fí­sica, Universidade Federal de Goiás, 74.001-970, Goiânia
- GO, Brazil}
\address{Departamento de Fí­sica, Universidade Federal de São Carlos, 13.565-905,
São Carlos - SP, Brazil}
\author{Norton G. de Almeida}
\address{Instituto de Fí­sica, Universidade Federal de Goiás, 74.001-970, Goiânia
- GO, Brazil}
\begin{abstract}
Recently, much emphasis has been given to genuinely quantum reservoirs
generically called fermionic reservoirs. These reservoirs are characterized
by having finite levels, as opposed to bosonic reservoirs,
which have infinite levels that can be populated via an increase in
temperature. Given this, some studies are being carried out to explore
the advantages of using quantum reservoirs, in particular in the operation
of heat machines. In this work, we make a comparative study of a thermal
refrigerator operating in the presence of either a bosonic or a fermionic
reservoir, and we show that fermionic reservoirs have advantages over
bosonic ones. We propose an explanation for the origin of these advantages by analyzing both the asymptotic behavior of the states of the qubits and the exchange rates between these qubits and
their respective reservoirs.
\end{abstract}
\maketitle

\section{Introduction}

With the development of quantum thermodynamics \citep{Gammer2004,Binder2019,Strasberg2021},
there is an increasing interest in the realization of thermal devices
operating at the quantum limit \citep{Abah2012,Alicki2014,Alecce2015,RoBnagel2016,Camati2019,Erdman2019,Chen2019}.
In particular, heat engines whose working substance consists of systems
with finite levels of energy, such as two-level systems, can absorb
from or deliver to their surroundings quantities of energy as small
as their corresponding energy gaps. In contrast, bosonic working substances,
modeled by quantum harmonic oscillators, have infinitely many levels,
where higher and higher levels can be populated by increasing temperature.
Comparative studies exploring the difference between a bosonic and
a fermionic working substance demonstrate that there are advantages
in considering working substances with finite levels for quantum engines
\citep{Henrich2007,Assis2020,Mendes2021,Mohammed2022}. In particular, systems
with finite energy levels, such as two-level systems, can exhibit
stationary states with population inversion \citep{Carr2013,Braun2013,Assis2019-1},
giving rise to absolute negative temperatures \citep{Abraham2017,Struchtrup2018}. The population
inversion associated with negative temperatures of the system requires
thermal reservoirs built with fermionic substances, as experimentally
demonstrated in \citep{Carr2013,Braun2013,Mendonca2020}. This inverted
population effect has been explored in some works \citep{Landsberg1980,Xi2017,Mendonca2020},
with remarkable impact on the efficiency of heat engines, as experimentally
shown in Ref. \citep{Assis2019-1,Mendonca2020}. On the other hand,
fermionic reservoirs \citep{Artacho1993,Henrich2007,Alvarez2010,Linden2010,Li2011,NuBeler2020,DelRe2020,Mikhailov2020}
built with two-level substances whose energy gap is $E$ does not
necessarily need to present population inversion, in which case its
temperature $T$ remains positive, with the average excitation number
given by the Fermi-Dirac distribution $n=1/\left(e^{E/T}+1\right)<0.5$,
in contrast with bosonic substances where the average excitation number
is given by the Bose-Einstein distribution $n=1/\left(e^{E/T}-1\right)$.
In the condition of positive temperature, one could imagine that there
would be no gain in considering fermionic reservoirs. However, in
this work we consider a study of case in which a refrigerator built
with two-level substances can present advantages when operating in
a fermionic environment as compared to a bosonic one, with both environments at positive temperatures. As the operating conditions are
kept the same for both environments, our results emphasize that the
presented advantage stems from the quantum nature of the fermionic
reservoir.

\section{Model}

\begin{figure}
\begin{centering}
\includegraphics[scale=0.95]{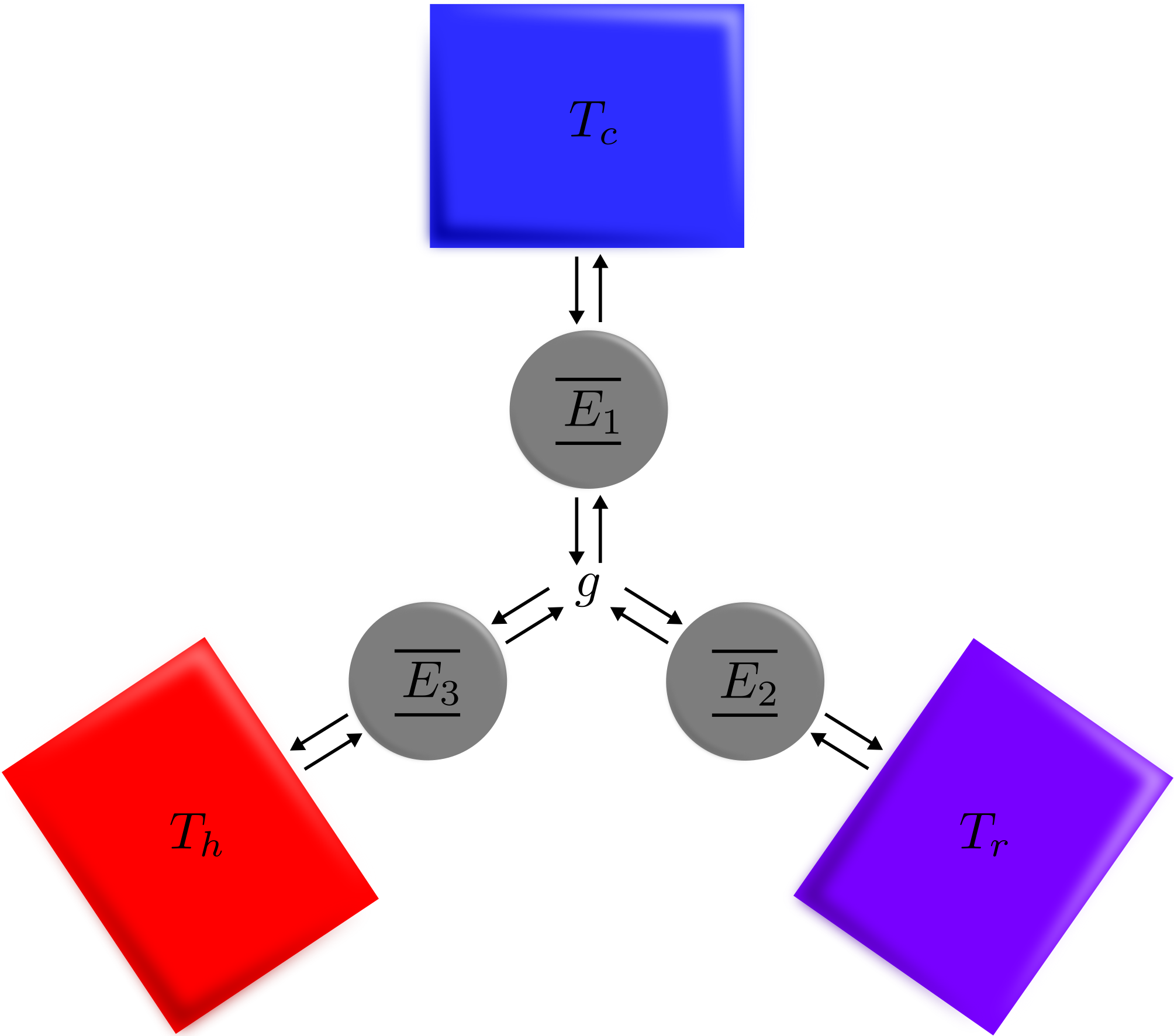}
\par\end{centering}
\caption{\label{fig:1}Schematic representation of the SCQR refrigerator and
its respective thermal reservoirs. The SCQR is composed of three interacting qubits having energy gaps $E_{1}, E_{2}$, and $E_{3}$ in contact with their respective reservoirs. Here, $T_{h}$ is the temperature of the hot reservoir, $T_{r}$ is the temperature of the "room" reservoir, $T_{c}$ is the temperature of the cold reservoir, and g is the coupling constant between the qubits. }

\end{figure}

In the present work, we consider a
self-contained quantum refrigerator (SCQR) composed of three
interacting qubits, each in contact with a specific thermal reservoir.
This SCQR was first proposed in Ref. \citep{Linden2010}, in which
the authors took into account only bosonic reservoirs. Recently, we
investigated this SCQR operating with one of the reservoirs being
a fermionic one at a negative temperature, see Ref. \citep{Damas2022}.
Here, as in Ref. \citep{Linden2010}, we approach the case in which
qubits 1, 2, and 3 interact respectively with a thermal reservoir
at a cold temperature $T_{c}>0$, a thermal reservoir at a \textquotedbl room\textquotedbl{}
temperature $T_{r}>0$, and a thermal reservoir at a hot temperature
$T_{h}>0$ - see the schematic shown in Fig. \ref{fig:1}. The device
in question works like a refrigerator when $T_{1}-T_{c}<0$, where
$T_{1}$ is the temperature of qubit 1. In this case, therefore, heat
flows from the cold reservoir to qubit 1. However, considering the
asymptotic state, this only occurs if the relations $E_{3}=E_{2}-E_{1}$,
with $E_{k}$ being the energy gap of qubit $k$ ($k=1,2,3$), and
$T_{c}<T_{r}<T_{h}$ are satisfied \citep{Linden2010}.

We assume the weak coupling limit and the Markovian regime governing
the dynamics of the SCQR, such that the master equation is \citep{Breuer2002,Li2011}
\begin{multline}
\frac{d\rho}{dt}=-i\left[H_{0}+H_{int}\right]+\\
\sum_{k=1}^{3}\Gamma_{B\left(F\right),k}^{\downarrow}\left(\sigma_{-,k}\rho\sigma_{+,k}-\frac{1}{2}\left\{ \sigma_{+,k}\sigma_{-,k},\rho\right\} \right)+\\
\sum_{k=1}^{3}\Gamma_{B\left(F\right),k}^{\uparrow}\left(\sigma_{+,k}\rho\sigma_{-,k}-\frac{1}{2}\left\{ \sigma_{-,k}\sigma_{+,k},\rho\right\} \right).\label{eq:1}
\end{multline}
Here, the free qubits Hamiltonian $H_{0}$ and the three-body
interaction Hamiltonian $H_{int}$ are given by
\begin{equation}
H_{0}=\frac{1}{2}E_{1}\sigma_{z,1}+\frac{1}{2}E_{3}\sigma_{z,3}+\frac{1}{2}E_{3}\sigma_{z,3}
\end{equation}
and 
\begin{equation}
H_{int}=g\left(\sigma_{-,1}\sigma_{+,2}\sigma_{-,3}+\sigma_{+,1}\sigma_{-,2}\sigma_{+,3}\right),
\end{equation}
where $\sigma_{z,k}$ is the z Pauli operator for qubit $k$,
$g$ is the coupling constant, and $\sigma_{-,k}$ ($\sigma_{+,k}$)
is the lowering (raising) Pauli operator for qubit $k$. Note that
Eq. \eqref{eq:1} governs the dynamics of either bosonic \citep{Breuer2002}
and fermionic \citep{Artacho1993,Alvarez2010,Li2011} thermal reservoirs:
if qubit $k$ is interacting with a bosonic (fermionic) thermal reservoir,
$\Gamma_{B,k}^{\downarrow}=\gamma_{k}\left(1+n_{B,k}\right)$ ($\Gamma_{F,k}^{\downarrow}=\gamma_{k}\left(1-n_{F,k}\right)$)
and $\Gamma_{B,k}^{\uparrow}=\gamma_{k}n_{B,k}$ ($\Gamma_{F,k}^{\uparrow}=\gamma_{k}n_{F,k}$),
where $\gamma_{k}$ is the dissipation rate and $n_{B,k}=1/(\text{e}^{E_{k}/T_{\phi_{k}}}-1)$
($n_{F,k}=1/(\text{e}^{E_{k}/T_{\phi_{k}}}+1)$) is the average excitation
number, being $\phi_{1}=c$, $\phi_{2}=r$, and $\phi_{3}=h$. Note that since for bosons $\Gamma_{B,k}^{\downarrow}=\gamma_{k}\left(1+n_{B,k}\right)$, $\Gamma_{B,k}^{\uparrow}=\gamma_{k}n_{B,k}$ and for fermions the average excitation number is limited  to 0.5 for positive temperatures, then $\Gamma_{B,k}^{\downarrow}$ and $\Gamma_{B,k}^{\uparrow}$ is always greater than $ \Gamma_{F,k}^{\downarrow} $ and $\Gamma_{F,k}^{\uparrow}$. To
obtain the asymptotic state of Eq. \eqref{eq:1} we used the quantum
optics toolbox \citep{Johansson2012,Johansson2013}.

\section{Results}

To compare the SCQR operating in the different configurations
involving bosonic and fermionic reservoirs, we start by fixing the
energies $E_{1}=1$, $E_{2}=5$, and $E_{3}=4$; the temperatures
$T_{c}=1,1.5,2$, and $T_{r}=2$; the coupling constant $g=10^{-2}$;
and the dissipation rates $\gamma_{1}=\gamma_{2}=\gamma_{3}=g$. Next,
we let $T_{h}$ vary from $10^{-1}$ to $10^{3}$. 

\begin{figure}[t]
\begin{centering}
\includegraphics{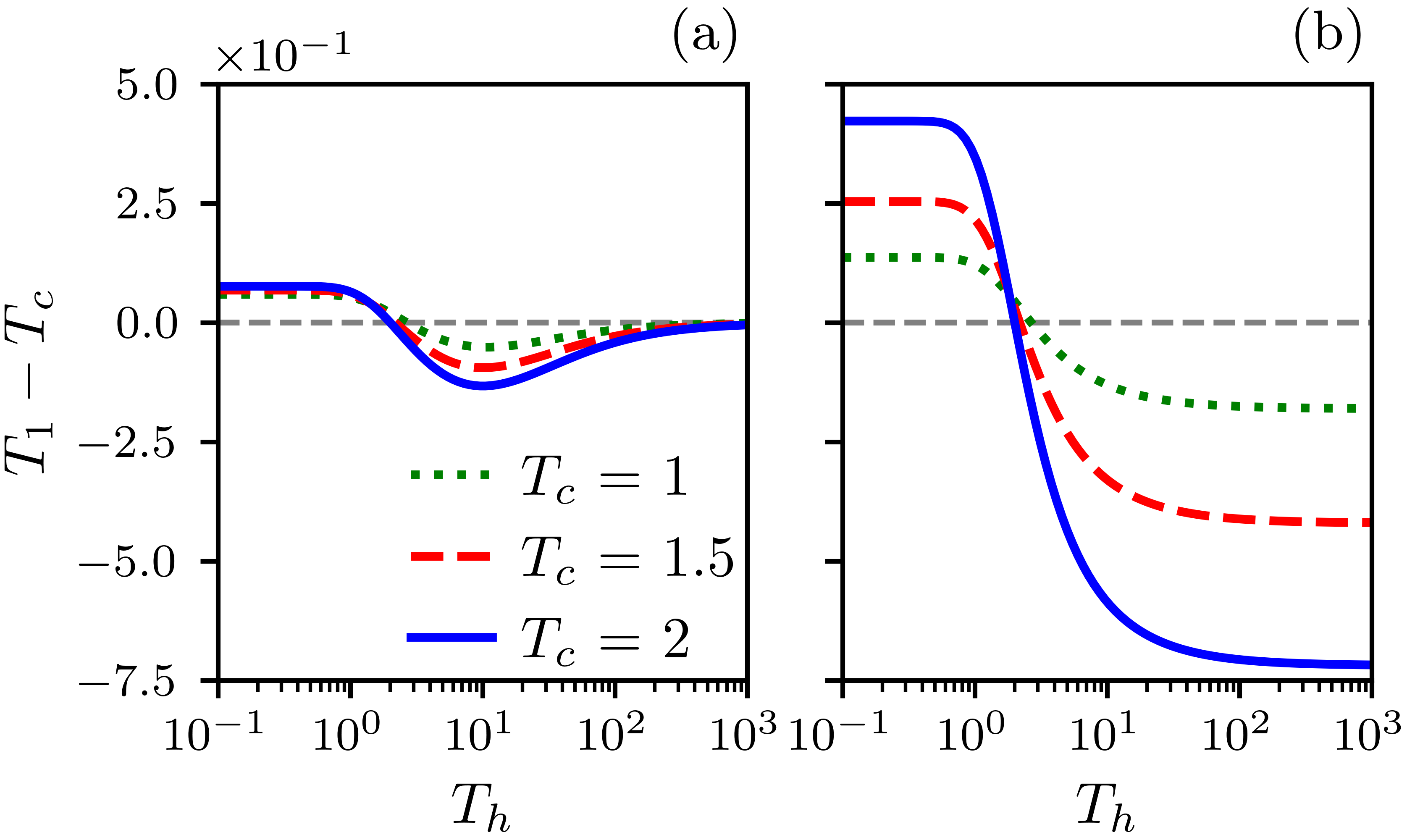}
\par\end{centering}
\caption{\label{fig:2}Temperature difference $T_{1}-T_{c}$ versus $T_{h}$
for (a) three bosonic and (b) three fermionic reservoirs, considering
three different values of $T_{c}$: $T_{c}=1$ (dotted green line),
$T_{c}=1.5$ (dashed red line) and $T_{c}=2$ (solid blue line). Refrigeration
occurs for $T_{1}-T_{c}<0$. Note the difference in behavior in the two figures: while in (a) temperature $T_{1}$ reaches a minimum and then starts to increase, in (b) $T_{1}$ decreases monotonically, practically stabilizing for sufficiently high $T_{h}$, thus indicating that
the lowest temperatures reached by qubit 1 occur for fermionic reservoirs.}
\end{figure}

Fig. \ref{fig:2}(a) shows the temperature difference $T_{1}-T_{c}$
versus $T_{h}$ (on logarithmic scale) for the SCQR working in a
bosonic environment for the cold temperatures $T_{c}=1$
(dotted green line), $T_{c}=1.5$ (dashed red line), and $T_{c}=2$
(solid blue line). As said before, cooling occurs when $T_{1}-T_{c}<0$.
Similarly, Fig. \ref{fig:2}(b) shows $T_{1}-T_{c}$ as a function
of $T_{h}$ for the same cold temperatures, but now the SCQR is surrounded
by fermionic reservoirs. In Fig. \ref{fig:2}(a), $T_{1}-T_{c}$ decreases
to a minimum value and then increases until it stabilizes at a negative
value close to zero, while, in Fig. \ref{fig:2}(b), $T_{1}-T_{c}$
stabilizes at its minimum value. Thus, the SCQR in the fermionic environment
has an advantage over the bosonic one, as its efficiency in cooling
qubit 1 does not decrease at higher values of $T_{h}$. Furthermore,
under fermionic reservoirs, qubit 1 reaches lower minimum
temperature values than when under bosonic reservoirs, as can be
seen from the difference $T_{1}-T_{c}$, which is more negative for
the fermionic environment (compare Figs. \ref{fig:2}(a) and
\ref{fig:2}(b)). 

According to our numerical simulations, when considering the
bosonic environment, the minimum values for $T_{1}$ are
$T_{1}=0.95$ (when $T_{c}=1$), $T_{1}=1.41$ (when $T_{c}=1.5$),
and $T_{1}=1.87$ (when $T_{c}=2$). On the other hand, when considering
three fermionic reservoirs, since the values for  $T_1$ continue to decrease with increasing  $T_h$, we take the minimum value for  $T_1$ when  $T_h =100$. These minimum values are $T_{1}=0.82$
(when $T_{c}=1$), $T_{1}=1.09$ (when $T_{c}=1.5$), and $T_{1}=1.29$
(when $T_{c}=2$). By considering $T_{c}$ as a reference we can then calculate the cooling percentage
 $\left(\left|T_{1}-T_{c}\right|/T_{c}\right)\times 100$
to compare how much the fermionic and bosonic reservoirs cools qubit
1 - see Tab. \ref{tab:1}, where 3B (3F) stands for three bosonic
(fermionic) reservoirs. 
\begin{table}
\begin{centering}
\begin{tabular}{cccc}
\toprule 
\multirow{2}{*}{$T_{c}$} &  & \multicolumn{2}{c}{(\%)}\tabularnewline
\cmidrule{2-4} \cmidrule{3-4} \cmidrule{4-4} 
 &  & 3B & 3F\tabularnewline
\midrule
\midrule 
1 &  & 5.14 & 17.57\tabularnewline
\midrule 
1.5 &  & 6.31 & 27.44\tabularnewline
\midrule 
2 &  & 6.65 & 35.32\tabularnewline
\bottomrule
\end{tabular}
\caption{\label{tab:1}Comparative values in the cooling percentage \textbf{$\left(\left|T_{1}-T_{c}\right|/T_{c}\right) \times 100$}
for the refrigerator working at three bosonic (3B) and three fermionic
(3F) reservoirs for three values of the reference temperature $T_{c}$.}
\par\end{centering}
\end{table}

Tab. \ref{tab:1} shows a significant difference in the cooling
percentage for the two sets of reservoirs: it is always higher when
using three fermionic reservoirs, thus clearly showing that the fermionic
environment is far more efficient in decreasing the temperature
$T_{1}$ than the bosonic one. Also, this percentage is better the
higher the reference temperature $T_{c}$, such that for $T_{c}=2$,
it can reach up to more than four times the value reached using only
bosonic reservoirs. For lower values of the cold temperatures $T_{c}$,
the percentage difference decreases but using fermionic reservoirs,
the cooling when $T_{c}=1$ is still more than double that of the
case of bosonic reservoirs alone. It is worth mentioning that by fixing
the SCQR parameters as we did, there is a limit to cooling
qubit 1. As we found numerically, the corresponding lowest cooling percentage
reached by qubit 1, irrespective of the type of reservoir used, occurs when $T_{c}\sim 0.48$. For temperatures lower than $T_{c}\sim 0.48$,  $T_{1}-T_{c}>0$,
meaning that the SCQR no longer works. Also, the percentage of cooling
decreases more and more as $T_{c}$ approaches $0.48$ for both
reservoirs. However, the percentage of cooling when using fermionic
reservoirs remains higher, as shown in Tab. \ref{tab:2}.

\begin{table}
\begin{centering}
\begin{tabular}{cccc}
\toprule 
\multirow{2}{*}{$T_{c}$} &  & \multicolumn{2}{c}{(\%)}\tabularnewline
\cmidrule{2-4} \cmidrule{3-4} \cmidrule{4-4} 
 &  & 3B & 3F\tabularnewline
\midrule
\midrule 
0.48 &  & 0.87 & 3.31\tabularnewline
\midrule 
0.60 &  & 2.20 & 7.39\tabularnewline
\midrule 
0.80 &  & 4.14 & 12.85\tabularnewline
\bottomrule
\end{tabular}\caption{\label{tab:2}Comparative values show that the cooling percentages
\textbf{$\left(\left|T_{1}-T_{c}\right|/T_{c}\right)\times 100$} for
fixed parameters decrease as the reference temperature $T_{c}$ approaches
$0.48$, where $T_{1}-T_{c}>0$. Note that even so fermionic reservoirs
are always more effective for cooling.}
\par\end{centering}
\end{table}

As we have seen, for fixed parameters we cannot cool down qubit 1 to zero
absolute. However, there is a strategy to keep up cooling toward zero
absolute, which is to isolate qubit 1 from its environment. This condition,
obtained by imposing $\gamma_{1}\rightarrow0$ or equivalently $\Gamma_{B\left(F\right),1}^{\downarrow}\rightarrow0$
and $\Gamma_{B\left(F\right),1}^{\uparrow}\rightarrow0$ in \eqref{eq:1},
allows us to obtain the following analytical solution for the temperature
of qubit 1: 
\begin{equation}
T_{1}=\frac{T_{c}}{1+\frac{E_{3}}{E_{1}}\left(1-\frac{T_{c}}{T_{h}}\right)},
\end{equation}
from which we can see that, if we let $E_{3}/E_{1}\rightarrow\infty$,
then $T_{1}\rightarrow0$. This result, obtained in Ref. \citep{Linden2010},
shows that there is no fundamental limit to cool down to zero absolute,
provided we can perfectly isolate qubit 1.

So far we have considered fermionic reservoirs for all qubits in the
SCQR. Other possibilities include the cases of combinations of bosonic
and fermionic reservoirs. In fact, considering the fermionic reservoir
as a quantum resource, it may be interesting to consider cases where
only one or two fermionic reservoirs are used. For this, it is necessary to
consider which qubit the fermionic reservoir is associated with. Let us use
a notation in which B (F) denotes the bosonic (fermionic) reservoir
and the order in which it appears in the sequence indicates which
qubit that reservoir is attached to. For example, the sequence BFB
indicates that qubit 1 is subjected to a bosonic reservoir, qubit
2 to a fermionic reservoir, and the third qubit to a bosonic reservoir.
Next, we investigate all configurations numerically and grouped the results
in Tab. \ref{tab:3}, ordering from highest to lowest percentage of
cooling and following the same procedure as in the previous tables, i.e., we took the minimum value for $T_{1}$.

\begin{table}
\begin{centering}
\begin{tabular}{cccccccccc}
\toprule 
\multirow{2}{*}{$T_{c}$} &  & \multicolumn{8}{c}{(\%)}\tabularnewline
\cmidrule{2-10} \cmidrule{3-10} \cmidrule{4-10} \cmidrule{5-10} \cmidrule{6-10} \cmidrule{7-10} \cmidrule{8-10} \cmidrule{9-10} \cmidrule{10-10} 
 &  & FBF & FFF & FBB & FFB & BBF & BFF & BBB & BFB\tabularnewline
\midrule
\midrule 
0.48 &  & 3.38 & 3.31 & 1.11 & 1.10 & 2.71 & 2.65 & 0.87 & 0.86\tabularnewline
\midrule 
0.80 &  & 13.09 & 12.85 & 7.31 & 7.22 & 7.76 & 7.62 & 4.14 & 4.09\tabularnewline
\midrule 
1 &  & 17.87 & 17.57 & 10.81 & 10.67 & 9.03 & 8.87 & 5.14 & 5.09\tabularnewline
\midrule 
1.5 &  & 27.28 & 27.44 & 18.65 & 18.43 & 10.28 & 10.13 & 6.31 & 6.24\tabularnewline
\midrule 
2 &  & 35.76 & 35.32 & 25.36 & 25.09 & 10.46 & 10.33 & 6.65 & 6.58\tabularnewline
\bottomrule
\end{tabular}
\par\end{centering}
\caption{\label{tab:3}Comparative values showing the cooling percentages $\left(\left|T_{1}-T_{c}\right|/T_{c}\right) \times 100$ for
several reservoir configurations. Here, for example, BFF means qubit
1 bound to a bosonic reservoir and qubits 2 and 3 bound to fermionic
reservoirs. Note that the highest percentage of cooling occurs for
FBF configuration, meaning that qubit 1 is bound to a fermionic reservoir,
qubit 2 is bound to a bosonic reservoir and qubit 3 is bound to another
fermionic reservoir.}

\end{table}

Interestingly, and c
ontrary to what one might think, the best case
does not occur when three fermionic reservoirs are used. As Tab. \ref{tab:3}
shows, the greatest cooling range occurs for FBF case, i.e., when
only qubit 1 and 3 are bound to fermionic reservoirs. Although the difference between the FBF and FFF configurations is small, it is still notable that the cooling percentage is higher when only two fermionic reservoirs are used instead of three.
\begin{table}[H]
\fontsize{5.5}{6}\selectfont
\begin{tabular}{ccccccccc}
\hline
                                          & FBF     & FFF     & FBB     & FFB     & BBF     & BFF     & BBB     & BFB     \\ \hline
$\Gamma_{B\left(F\right),1}^{\downarrow}$ & 0.00622 & 0.00622 & 0.00622 & 0.00622 & 0.02541 & 0.02541 & 0.02541 & 0.02541 \\ \hline
$\Gamma_{B\left(F\right),1}^{\uparrow}$   & 0.00378 & 0.00378 & 0.00378 & 0.00378 & 0.01541 & 0.01541 & 0.01541 & 0.01541 \\ \hline
$\Gamma_{B\left(F\right),2}^{\downarrow}$ & 0.01089 & 0.00924 & 0.01089 & 0.00924 & 0.01089 & 0.00924 & 0.01089 & 0.00924 \\ \hline
$\Gamma_{B\left(F\right),2}^{\uparrow}$   & 0.00089 & 0.00076 & 0.00089 & 0.00076 & 0.00089 & 0.00076 & 0.00089 & 0.00076 \\ \hline
$\Gamma_{B\left(F\right),3}^{\downarrow}$ & 0.00510 & 0.00510 & 0.02812 & 0.02812 & 0.00510 & 0.00510 & 0.02812 & 0.02812 \\ \hline
$\Gamma_{B\left(F\right),3}^{\uparrow}$   & 0.00490 & 0.00490 & 0.01812 & 0.01812 & 0.00490 & 0.00490 & 0.01812 & 0.01812 \\ \hline
\end{tabular}\caption{\label{Tab 4}Exchange rates $\Gamma_{B\left(F\right),k}^{\downarrow}$ and $\Gamma_{B\left(F\right),k}^{\uparrow}$ of the
qubit k with their respective reservoirs. From this Table, we see
that the lowest (highest) exchange rates occur for fermionic (bosonic)
reservoirs. Note that the FBF configuration presents the smallest exchange rates for qubit 1. This explains why it is  more effective
for cooling qubit 1 - see main text. The temperatures used are $T_{c}=2$, $T_{r}=2$, and $T_{h}$ varies to minimize the values of $T_{1}$, as the SCQR behavior changes as shown in Figs. \ref{fig:2}(a) and
\ref{fig:2}(b). The same pattern occurs if we use other values for $T_{c}$.}
\end{table}

In this regard, note that the FFB and BFF sequences, although each also contain only two fermionic reservoirs and one bosonic reservoir, they have a lower cooling percentage than that of the FBF sequence. For instance,
for $T_{c}=2$, the cooling percentage of FBF is $29.72\%$, which is higher
than that for sequence FFB ($25.09\%$) and BFF ($10.33\%$). The explanation for this fact is given below. Remembering that the
lowest temperature for qubit 1, which is the qubit we want to cool,
occurs when it is completely isolated, it is to be expected, therefore,
that when the exchange rates $\Gamma_{F(B),1}^{\uparrow}$ and $\Gamma_{F(B),1}^{\downarrow}$
of qubit 1 with its reservoir are the lowest possible, the cooling
will take place more effectively, with perfect insulation being the
best case. In Tab. \ref{Tab 4} we
show the exchange rates $\Gamma_{F(B),k}^{\uparrow}$ and $\Gamma_{F(B),k}^{\downarrow}$, $k=1,2,3$, for the
k-th qubit for temperatures $T_{c}=2$, $T_{r}=2$, and $T_{h}=10$. From Tab. \ref{Tab 4} we see that the $\Gamma_{F(B),1}^{\uparrow}$ and $\Gamma_{F(B),1}^{\downarrow}$ rates are the smallest whenever the sequence starts with F,
and remain the smallest irrespective of the temperatures
used, according to our simulations. Another relevant point to be considered, which we have already shown in Figs. \ref{fig:2}(a) is that bosonic reservoirs have the disadvantage of making the cooling non-monotonic, and therefore less effective. Thus, obtaining better cooling percentages requires that the last reservoir be fermionic, as we also verified in our numerical simulations. The role of the nature of the second reservoir and its relevance to cooling percentages is quite complex. In our numerical simulations, we were able to identify that the best cooling percentages occur, whenever one bosonic reservoir and two fermionic reservoirs are used, in the sequence FBF - see Tab. \ref{tab:3}. Regarding other configurations with lower cooling percentages but yet involving two fermionic and one bosonic reservoir, note for example that the FFB sequence may have higher or lower cooling percentages than the BFF sequence depending on whether the temperature $T_c$ is higher or lower than unity.

\section{Conclusion}

Recent studies on heat machines have used quantum reservoirs as a
resource to obtain better performances both in engines and in refrigerators
\citep{Landsberg1980,Xi2017,Assis2019-2,Assis2020}. For example,
fermionic reservoirs have been explored in previous works, especially
in their purely quantum characteristic of presenting population inversion
\citep{Assis2019-1,Mendonca2020}, which, in turn, is associated with
negative effective temperatures \citep{Abraham2017,Strasberg2021}.
Here we explore the quantum nature of fermionic reservoirs without
taking population inversion into account, such that we restrict to
the domain of positive temperatures. Using a qubit-based refrigerator
model proposed in Ref. \citep{Linden2010}, we show that, once the
operating parameters of the refrigerator are fixed, the use of fermionic
reservoirs allows to obtain better results, with respect to the cooling
capacity, than the use of bosonic reservoirs. We have verified, for
example, that when the qubit to be cooled cannot be perfectly insulated,
the use of only fermionic reservoirs allows to reach lower temperatures
than the use of only bosonic reservoirs. In addition, contrary to
what might be thought, the cooling can be more effective, in the sense
of obtaining a higher percentage of cooling, when instead of three,
only two fermionic reservoirs are used. We show that an explanation of this somewhat
unexpected result is due to the exchanged rates between qubit 1 and its reservoir as well as to the behavior of the asymptotic cooling of qubit 1 when subjected to different types of reservoirs. In summary, when the condition for perfect
insulation cannot be reached, our results unequivocally demonstrate
the superiority of the fermionic reservoir in the process of cooling
qubits to the lowest possible temperatures.
\begin{acknowledgments}
We acknowledge financial support from the Brazilian agencies: Coordenação
de Aperfeiçoamento de Pessoal de Nível Superior (CAPES), financial
code 001, National Council for Scientific and Technological Development
(CNPq), grant 311612/2021-0 and 301500/2018-5, São Paulo Research
Foundation (FAPESP), grant 2021/04672-0, and Goiás State Research
Support Foundation (FAPEG). This work was performed as part of the
Brazilian National Institute of Science and Technology (INCT) for
Quantum Information, grant 465469/2014-0.
\end{acknowledgments}

\bibliographystyle{apsrev4-1}
\bibliography{References}

\end{document}